\documentclass[fleqn,twoside]{article}
\usepackage{espcrc2,epsfig}

\newcommand{\msbar}{{\rm \overline{MS\kern-0.05em}\kern0.05em}}

\newcommand{\AmS}{{\protect\the\textfont2
  A\kern-.1667em\lower.5ex\hbox{M}\kern-.125emS}}

\newcommand{\gsim}{{\protect
  \kern.18em\lower.5ex\hbox{$\stackrel{>}{\sim}$}\kern.25em}}
\newcommand{\lsim}{{\protect
  \kern.17em\lower.5ex\hbox{$\stackrel{<}{\sim}$}\kern.23em}}

\newcommand\be{\begin{equation}}
\newcommand\ee{\end{equation}}
\newcommand\bea{\begin{eqnarray}}
\newcommand\eea{\end{eqnarray}}

\newcommand{\mps}{M_{\rm P}}
\newcommand{\fps}{f_{\rm P}}

\def\mov{m_{\rm ov}}
\def\mstr{m_{\rm s}}
\def\mq{m_{\rm q}}
\def\mw{m_{\rm w}}
\def\mbar{\kern1pt\overline{\kern-1pt m\kern-1pt}\kern1pt}
\def\mbarSF{{\kern1pt\overline{\kern-1pt m\kern-1pt}\kern1pt}_{\rm SF}}

\def\mev{\rm MeV}
\def\gev{\rm GeV}
\def\fm{\rm fm}

\def\xref{x_{\rm ref}}
\def\UM{U_{\rm M}}
\def\UP{U_{\rm P}}

\def\za{Z_{\rm A}}

\def\zp{Z_{\rm P}}
\def\zs{Z_{\rm S}}
\def\zshat{\widehat{Z}_{\rm S}}

\def\zm{Z_{\rm m}}
\def\zM{Z_{\rm M}}

\def\bp{b_{\rm P}}

\title{%
\vspace{-5.5cm}
\begin{flushright}
       {\normalsize CERN-TH/2001--287}    \\[-0.2cm]
       {\normalsize CPT-2001/PE.4252}\\[-0.2cm]
       {\normalsize DESY 01--174}\\[-0.2cm]
       {\normalsize IFIC/01--58}\\[-0.2cm]
       {\normalsize INT-ACK 01--20}\\[-0.2cm]
       {\normalsize FTUV-011025}\\[-0.2cm]
       {\normalsize LTH\,525}
\end{flushright}
       \vspace{1.7cm}
Scalar condensate and light quark masses from overlap fermions
      }

\author{
        Pilar Hern\'andez\address{CERN, Theory Division,                     
        1211 Geneva 23, Switzerland}
       \thanks{On leave of absence from Departamento de F\'{\i}sica
        Te\'orica, Universidad de Valencia, Spain},
        Karl Jansen \address{NIC/DESY Zeuthen,                                          Platanenallee 6, D-15738 Zeuthen, Germany}
       \thanks{Presented by K. Jansen and H. Wittig at {\sl Lattice 2001}},  
        Laurent Lellouch\address{Centre de Physique Th\'eorique,
        Case 907, CNRS Luminy, F-13288 Marseille, France}                      
        and Hartmut Wittig\address{Division of Theoretical Physics,
        Department of Mathematical Sciences, University of Liverpool,
        UK}
        \address{DESY,                                   
        Notkestr. 85, D-22603 Hamburg, Germany}\,\,$^\dagger$
    }
           
\begin{document}

\begin{abstract}
We have studied pseudoscalar correlation functions computed using the
overlap operator. Within the accuracy of our calculation we find that
the quark mass dependence agrees with the prediction of lowest-order
Chiral Perturbation Theory $\chi$PT for quark masses in the range of
$m\sim\mstr/2-2\mstr$. We present the results of an analysis which
assumes lowest-order $\chi$PT to be valid to extract the low-energy
constants $\Sigma$ and $\fps$, as well as the strange quark
mass. Non-perturbative renormalization is implemented via a matching
procedure with data obtained using Wilson fermions in the
Schr\"odinger functional set-up. We find that the scalar condensate
computed here agrees with the one obtained previously through a
finite-size scaling analysis.
\end{abstract}

\maketitle

\section{INTRODUCTION}

Testing for spontaneous chiral symmetry breaking in QCD by determining
whether a scalar condensate is developed and computing its size is a
challenge tailored for lattice field theory. One way to extract the
scalar condensate is to look at the behaviour of the
pseudoscalar mass $\mps$ as a function of the quark mass
$m$. Lowest-order chiral perturbation theory ($\chi$PT) predicts this
behaviour to be linear \cite{chir:Sha92,chir:BerGol92},
\begin{equation}
\left(a\mps\right)^2 = B_M am\;\; ,B_M=4a\Sigma/f_\chi^2
\label{mpiq}
\end{equation}
where $\Sigma$ and $f_\chi$ are the only constants of the lowest-order
effective chiral Lagrangian. In the full theory they coincide with the
quark condensate and the pseudoscalar decay constant in the chiral
limit, respectively.

Within the numerical accuracy of our calculation in the quenched
approximation (see below) it turns out that the linear behaviour of
eq.~(\ref{mpiq}) is indeed satisfied for a large range of quark
masses, extending from $m\sim\mstr/2$ to $m\sim2\mstr$. However, the
linearity of the data is actually surprising, since it is observed in
a range of quark masses where higher-order terms in the quenched
chiral Lagrangian could be relevant. Furthermore, at one-loop order
one expects logarithmic corrections to appear
\cite{chir:Sha92,chir:BerGol92}. We suspect that the observed
linearity is actually the result of a large cancellation between
various higher-order effects. A detailed discussion into the r\^ole of
these effects will be presented elsewhere \cite{cond:paperIV}. In the
preliminary analysis presented here, we restrict ourselves to
lowest-order $\chi$PT.

Our assumption that lowest-order $\chi$PT gives a good description of
the mass behaviour receives support from the fact that we have a
completely independent method for determining $\Sigma$, which gives
consistent results. Indeed, using overlap fermions it was shown that
the bare subtracted scalar condensate could be obtained through
finite-size scaling methods \cite{cond:paperI}. This analysis has now
been repeated by other authors \cite{cond:DeGrand,cond:bern_lat01},
who have obtained consistent results. Higher-order effects might be
important for this method as well, though here the corrections will be
different. Again, we restrict ourselves to lowest-order quenched
$\chi$PT, which is consistent with the numerical data
\cite{cond:paperI}, being fully aware that this assumption must
eventually be checked.

In order to be useful for phenomenological applications, the bare
condensate has to be renormalized. Renormalization factors which
relate matrix elements of local composite operators in lattice
regularization to a standard continuum renormalization scheme such as
$\msbar$ can be computed in lattice perturbation theory. However, it
is well known that perturbation theory in the bare coupling $g_0$ does
not converge very well, and although the situation can be improved by
considering so-called ``mean-field improvement'' \cite{lepenzie93}, it
is preferable to determine these factors non-perturbatively. One
method to compute the renormalization factor for the scalar condensate
computed using the overlap operator has been described in a recent
publication~\cite{cond:paperIII} and will be briefly reviewed in the
following section.

\section{NON-PERTURBATIVE RENORMALIZATION}

The chiral Ward identities imply that in lattice regularizations that
preserve chiral symmetry the renormalization factor of the scalar
density is the inverse of that for the quark mass (see
e.g. \cite{chiral:AlFoPaVi})
\be
   \zp=\zs=\frac{1}{\zm}.
\label{eq_zszpzm}
\ee
The non-perturbative renormalization of quark masses via an
intermediate Schr\"odinger functional (SF) scheme has been studied by
ALPHA \cite{mbar:pap1}, who have computed the relations between the
running mass in the SF scheme, $\mbarSF$, and the renormalization
group invariant (RGI) quark mass~$M$, as well as the non-perturbative
matching coefficient between $\mbarSF$ and the bare current quark mass
$\mw$ for O($a$) improved Wilson fermions. Thus, in order to determine
$\zs$ for overlap fermions, it is sufficient to compute the ratio of
the bare mass in the overlap Lagrangian, $\mov$, and $\mbarSF$.
However, the formulation of the SF for overlap fermions is not
straightforward, owing to the fact that the SF boundary conditions are
incompatible with the Ginsparg-Wilson relation. We have therefore
employed the following, more indirect approach.

The relation between the bare mass $\mov(g_0)$ and the RGI quark
mass~$M$ is given by
\be
   M = \zM(g_0)\,\mov(g_0).
\ee
The ratio $M/\mov(g_0)$ can be rewritten as
\bea
  \frac{M}{\mov(g_0)} &=& \frac{M}{\mw(g_0^\prime)}\cdot
                       \frac{\mw(g_0^\prime)}{\mov(g_0)}   \\
                   &=& \zM^{\rm w}(g_0^\prime)\cdot
            \frac{(r_0\mw)(g_0^\prime)}{(r_0\mov)(g_0)},
\eea
where $g_0^\prime\not=g_0$ in general, and the hadronic radius $r_0$
\cite{pot:r0} is used to set the scale. The factor $\zM^{\rm w}$ has
been computed in \cite{mbar:pap1} for a wide range of couplings. The
ratio $M/\mov(g_0)$ is then obtained by determining $(r_0\mov)$ and
$(r_0\mw)$ at a reference value $\xref$ of some observable, say
$\xref=(r_0\mps)^2$. Furthermore, the combination $\zM^{\rm
w}(g_0^\prime)(r_0\mw)(g_0^\prime)$ is a renormalized, dimensionless
quantity. We can then define the universal factor $\UM$ in the
continuum limit as
\be
   \UM = \lim_{g_0^\prime\to0}\left\{\zM^{\rm w}(g_0^\prime)
   (r_0\,\mw)(g_0^\prime)\right\}
   \Big|_{(r_0\mps)^2=x_{\rm ref}.}
\label{eq_umdef}
\ee
Since the renormalization of the scalar condensate is the inverse of
that of the quark mass, we define the renormalization factor $\zshat$,
which relates the bare and RGI condensates as
\be
   \zshat(g_0) = \frac{(r_0\mov)}{U_{\rm M}
                 }\Big|_{(r_0\,\mps)^2=x_{\rm ref}.}
\label{eq_zsum}
\ee
At this point it is clear that all reference to the bare coupling
$g_0^\prime$ has disappeared, and the only part that retains a
dependence on the lattice regularization is $(r_0\mov)$.

The quantity $\UM$ is in fact the RGI quark mass corresponding to
$(r_0\mps)^2=\xref$. The results from ref.\,\cite{mbar:pap3} are then
easily converted into estimates for $\UM$ at several values of
$\xref$. We stress that no additional calculations are required to
determine the universal factor $\UM$, but that it can be derived
easily from existing results in the literature. Furthermore, using the
continuum value $\UM$ ensures that cutoff effects of O($a^2$)
associated with the intermediate use of Wilson fermions at bare
coupling $g_0^\prime$ are entirely removed from the renormalization
condition. As described in detail in \cite{cond:paperIII}, additional
cutoff effects in the case of non-zero $g_0^\prime$ can be quite
substantial. 

As indicated by eq.~(\ref{eq_zszpzm}), the renormalization of the
scalar condensate and the quark mass are inversely proportional if
chiral symmetry is preserved. If eq.~(\ref{eq_zsum}) is rewritten
as
\be
    \left.\UM\right|_{\xref} = \frac{1}{\zshat(g_0)}\,
    \left\{(r_0\mov)(g_0)\right\}_{\xref}
\ee
it becomes clear that this condition fixes $\zshat(g_0)$ by requiring
that the renormalized quark mass reproduces the continuum result for
Wilson fermions at {\em every} non-zero value of $g_0$. In other
words, since the quark mass $\mov$ is required to fix $\zshat$, the
latter cannot be used to obtain independent predictions for
renormalized quark masses for overlap fermions.

This problem can be circumvented after realizing that an alternative
renormalization condition for $\zshat$ is provided by the matrix
element of the pseudoscalar density. The derivation is entirely
analogous to the previous case. If we use the notation
\be
  G_{\rm P}=\langle0|P|\hbox{PS}\rangle
\ee
as a shorthand for the matrix element of the pseudoscalar density $P$,
and the superscripts ``RGI'', ``ov'' and ``w'' to distinguish the RGI
matrix element from the one in the overlap and O($a$) improved Wilson
regularizations, we find
\be
  \frac{G_{\rm P}^{\rm RGI}}{G_{\rm P}^{\rm ov}(g_0)} =
  \frac{G_{\rm P}^{\rm RGI}}{G_{\rm P}^{\rm w}(g_0^\prime)}
  \cdot
  \frac{(r_0^2G_{\rm P}^{\rm w})(g_0^\prime)}
       {(r_0^2G_{\rm P}^{\rm ov})(g_0)}.
\ee
The ratio $G_{\rm P}^{\rm RGI}/G_{\rm P}^{\rm w}(g_0^\prime)$ is given
by \cite{impr:pap1,mbar:pap1}
\be
  \frac{G_{\rm P}^{\rm RGI}}{G_{\rm P}^{\rm w}(g_0^\prime)}
  = \zp^{\rm w}(g_0^\prime,\mu_0)\frac{\mbarSF(\mu_0)}{M}
  (1+\bp{a\mq}), 
\ee
where $\mq$ is the bare subtracted quark mass for Wilson fermions. The
improvement coefficient $\bp$ is known in one-loop perturbation theory
\cite{impr:pap5}, and the combination $\zp^{\rm w}\mbarSF/M$ is known
from ref.\,\cite{mbar:pap1}. Again one can consider a reference point
$\xref$ and define a universal factor in the continuum limit by
\bea
  \UP= \lim_{g_0^\prime\to0} &&\hspace{-0.5cm}
  \left\{\zp^{\rm w}(g_0^\prime,\mu_0)
  \frac{\mbarSF}{M} \right. \nonumber \\
  &&\hspace{-2.0cm} \left.\left.\phantom{\frac{\mbar}{M}}\times
  {(1+\bp a\mq)(r_0^2 G_{\rm P}^{\rm w})(g_0^\prime)}\right\}
  \right|_{(r_0\mps)^2=\xref,} 
  \label{eq_updef}
\eea
such that the alternative definition for $\zshat(g_0)$ is obtained
from
\be
   \zshat(g_0) = \frac{\UP}{(r_0^2 G_{\rm P}^{\rm ov})(g_0)}
                 \Big|_{(r_0\,\mps)^2=x_{\rm ref}.}
\label{eq_zsup}
\ee
As in the previous case one can easily convert the results of
\cite{mbar:pap3} to obtain the continuum factor $\UP$.

Our results for $(r_0\mov)$ and $(r_0^2 G_{\rm P})$ at several values
of $\xref$ were extracted from the same pseudoscalar correlation
functions used to determine the condensate according to
eq.~(\ref{mpiq}). In Table~\ref{tab_ZShat_res} we list our estimates
for $\zshat$ determined at two values of the lattice spacing for
several reference points. As one can see, the results for the two
different renormalization conditions are entirely consistent within
errors. In the following we will use the results from
eq.~(\ref{eq_zsup}) for $\xref=3.0$. We also note that the
non-perturbative determination of $\zs$ in ref.~\cite{cond:boston},
which is based on a different technique \cite{renorm_mom:paper1},
agrees with our data at $\beta=6.0$.

\begin{table}[t]
\caption{Results for $\zshat$. \label{tab_ZShat_res}}
\begin{tabular}{cl l l}
\hline
\hline
        &  & eq.~(\ref{eq_zsum})  &  eq.~(\ref{eq_zsup})  \\
$\beta$ & $\xref$ & $\zshat$      &  $\zshat$      \\
\hline
5.85    & 1.5736  & 1.05(25)      &  1.03(15)      \\
        & 3.0     & 1.04(8)       &  1.04(8)       \\
        & 5.0     & 0.99(4)       &  0.99(6)       \\
\hline
6.00    & 1.5736  & 0.98(17)      &  1.05(14)      \\
        & 3.0     & 1.03(8)       &  1.07(8)       \\
        & 5.0     & 1.00(5)       &  1.03(6)       \\
\hline
\hline
\end{tabular}
\end{table}

\section{SCALAR CONDENSATE}

\begin{figure}[tb]
\vspace{-1.5cm}
\hspace{-0.5cm}
\centerline{ \epsfysize=7.8cm
             \epsfxsize=8.0cm
             \epsfbox{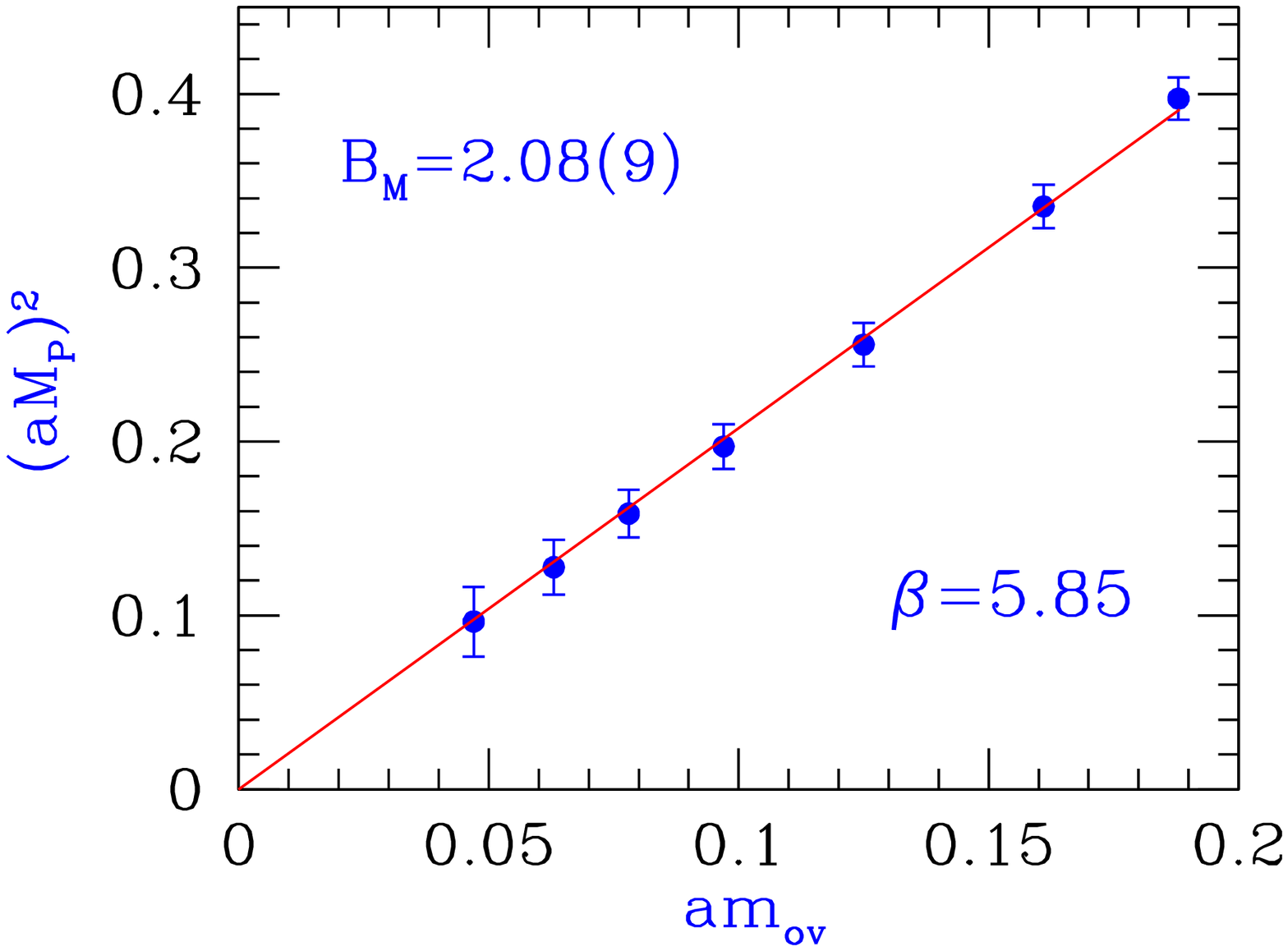}}
\vspace{-2.0cm}
\caption{The pseudoscalar mass $(a\mps)^2$  
as a function of the bare quark mass at $\beta=5.85$. \label{figure1}}
\end{figure}

\begin{figure}[tb]
\vspace{-1.5cm}
\hspace{-0.5cm}
\centerline{ \epsfysize=7.8cm
             \epsfxsize=8.0cm
             \epsfbox{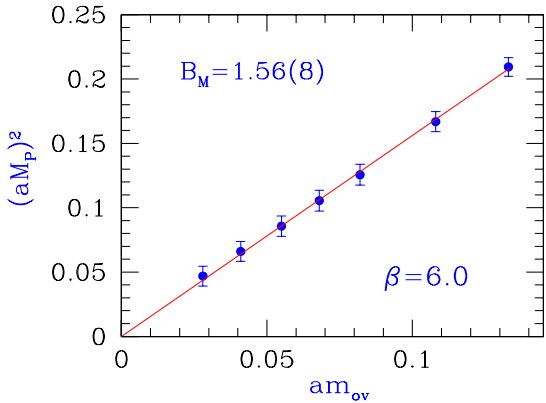}}
\vspace{-2.0cm}
\caption{The pseudoscalar mass $(a\mps)^2$           
as a function of the bare quark mass at $\beta=6.0$. \label{figure2}}
\end{figure}

We computed pseudoscalar propagators using overlap fermions
\cite{chiral:ovlp} at two values of the lattice spacing. The
parameters of the runs are listed in Table~\ref{runparameters}, and
further simulation details can be found in
\cite{cond:paperII,cond:paperIII}. Pseudoscalar meson masses were
extracted by fitting the numerical data obtained for the zero momentum
pseudoscalar propagator to a standard single cosh form. The masses
$\mps$ are shown as a function of the bare quark mass in
Figs.~\ref{figure1} and~\ref{figure2}. We observe that $\mps^2$ is
consistent with being a linear function of the quark mass in the whole
region.

\begin{table}[tpb]
\caption{Simulation parameters.\label{runparameters}}
\begin{tabular}{ccccc}
\hline 
\hline
 $\beta$ & $\rho$ & Vol. & cfgs & $a\mov$\\
\hline
 5.85 & 1.6 & $10^3\times 24$ & 50 & 0.047$-$0.188\\
 6.00 & 1.4 & $14^3\times 24$ & 48 & 0.028$-$0.133\\
\hline
\hline
\end{tabular}
\end{table}

Using eq.~(\ref{mpiq}) an estimate for $a\Sigma/f_\chi^2$ can be
extracted easily. Combining these results with the renormalization
factor $\zshat$ at the relevant lattice spacing and using $r_0/a$ from
\cite{pot:r0_SU3} to set the scale, we obtain for the RGI condensate
the following estimates:
\be
   \widehat{\Sigma}=\left\{\begin{array}{l}
           0.0141(6)(11)\,\gev^3,\quad\beta=5.85 \\
           0.0144(6)(11)\,\gev^3,\quad\beta=6.0.
                           \end{array} \right.
\label{eq_GMOR_ov}
\ee
Here we have used $f_\chi=128\,\mev$ and $r_0=0.5\,\fm$ to convert
into physical units. The first error is the statistical uncertainty in
the determination of $a\Sigma/f_\chi^2$, whereas the second is due to
the error in the renormalization factor. The results in
eq.~(\ref{eq_GMOR_ov}) can now be compared with the estimate obtained
from the finite-size scaling (FSS) analysis at $\beta=5.85$, which is
\be
   \widehat{\Sigma}= 0.0138(16)(10)\,\gev^3,\quad\beta=5.85.
\label{eq_FSS_ov}
\ee
This result is fully consistent with the above value. We note,
however, that the scale $r_0$ enters as $r_0^3$ in
eq.~(\ref{eq_FSS_ov}), and that the extraction of the FSS condensate
in physical units does not depend at all on the value of
$f_\chi$. Thus, uncertainties in the lattice scale affect the two
methods of extracting $\widehat\Sigma$ in different ways.

\begin{figure}[htb]
\centerline{ \epsfysize=7.8cm
             \epsfxsize=8.0cm
             \epsfbox{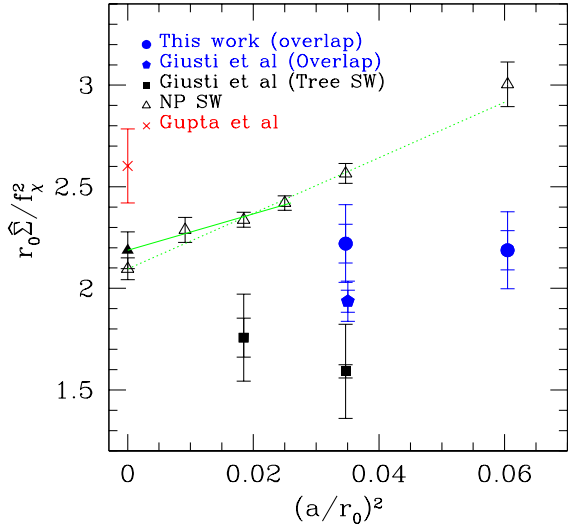}}
\vspace{-1.3cm}
\caption{Comparison of different computations of the scalar
condensate. We also show the result from Wilson fermions
\protect\cite{mbar:pap3} extrapolated to the continuum. The double
error bars denote the statistical and the quadratically combined
statistical and systematic uncertainties. \label{figure3}}
\vspace{-0.3cm}
\end{figure}

Of course, eq.~(\ref{mpiq}) can also be used to compute $\Sigma$ from
data obtained in simulations with O($a$) improved Wilson fermions,
despite the fact that chiral symmetry is explicitly broken in this
formulation. What makes this strategy attractive after all is the fact
that high statistics data at several values of $\beta$ are available,
allowing for a controlled continuum extrapolation. Combining our
Wilson data at $\beta=5.85$ with the data from the ALPHA Collaboration
\cite{mbar:pap3}, we have extrapolated $r_0\widehat{\Sigma}/f_\chi^2$
to the continuum limit as a function of $(a/r_0)^2$ (see
Fig.~\ref{figure3}). The continuum results are entirely consistent
with our results obtained using overlap fermions at the present level
of accuracy. Fig.~\ref{figure3} also shows a compilation of results
from various other determinations
\cite{quark:gupta,cond:APE,cond:boston}. It demonstrates that there is
not only broad agreement of the results but also that lattice spacing
effects with overlap fermions seem to be rather small.

\section{STRANGE QUARK MASS}

The behaviour of the dependence of $(a\mps)^2$ on the quark mass
$(am)$ also allows us to calculate the renormalization group invariant
strange quark mass $M_s$. Using again $r_0$ to set the scale and
$m_K=495\,\mev$ to fix the quark mass we find the results shown in
Fig.~4 for $M_s+\hat{M}$, where $\hat{M}$ is the average of the RGI up
and down quark masses. The agreement with the results obtained using
O$(a)$ improved Wilson fermions in the continuum limit is good,
although the error bars in the overlap case are rather large due to
our small statistics.

\begin{figure}[htb]
\centerline{ \epsfysize=7.8cm
             \epsfxsize=8.0cm
             \epsfbox{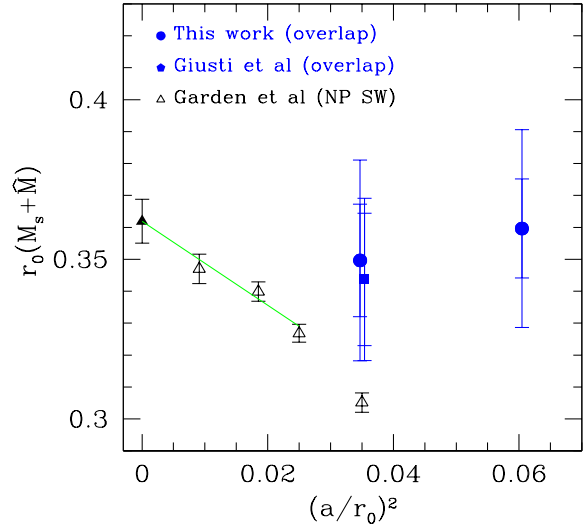}}
\vspace{-1.3cm}
\caption{The renormalization group invariant strange quark mass as a
function of the lattice spacing. \label{figure4}}
\vspace{-0.3cm}
\end{figure}

\section{DECAY CONSTANT}

\begin{figure}[htb]
\centerline{ \epsfysize=7.8cm
             \epsfxsize=8.0cm
             \epsfbox{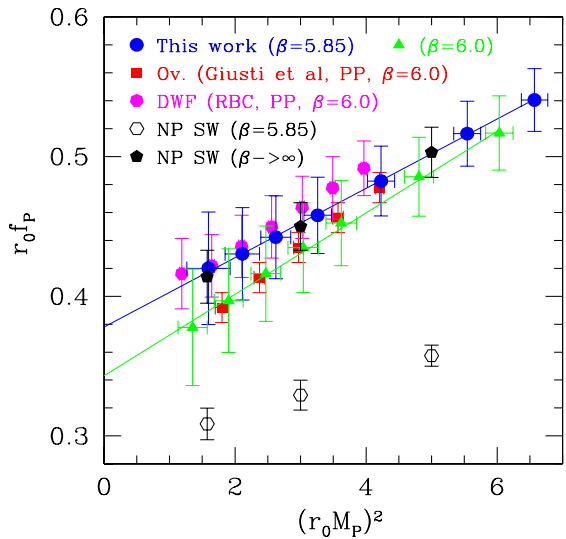}}
\vspace{-1.3cm}
\caption{$r_0\fps$ versus $(r_0\mps)^2$ for various
discretizations. The lines are fits to our data at $\beta=5.85$ (upper
line) and $\beta=6.0$ (lower line).\label{figure5}} 
\end{figure}

Since the overlap formalism preserves chiral symmetry at non-zero
lattice spacing, one can determine the pseudoscalar decay constant
$\fps$ without any further renormalization. By contrast, in
simulations using Wilson fermions the renormalization factor $\za$ of
the axial current must be computed.
In our work we have determined $\fps$ through the PCAC relation, i.e.
\be
   \fps = 2\frac{\mov}{\mps^2}\langle0|P|\rm PS\rangle,
\ee
where $\mov$ is an input parameter in the simulation, while $\mps$ and
$\langle0|P|\rm PS\rangle$ are extracted from pseudoscalar correlation
functions. In order to compare results from different fermionic
discretizations we have plotted $r_0\fps$ versus $(r_0\mps)^2$ in
Fig.~\ref{figure5}. One observes rather good agreement among the data
employing either overlap (this work, \cite{cond:boston}) or Domain
Wall \cite{dwf:RBC_spect} fermions, as well as with the Wilson results
extrapolated to the continuum limit with non-perturbative estimates
for $\za$ \cite{impr:pap4}. The fact that the overlap data at
$\beta=5.85$ are already close to the continuum Wilson results
suggests that discretization errors for this quantity may be small. By
contrast, O($a$) improved Wilson results at $\beta=5.85$ differ
substantially from their continuum limit. The apparent smallness of
cutoff effects in the overlap case is actually the reason for the
excellent agreement of $\zshat$ estimated either through
eq.~(\ref{eq_zsum}) or~(\ref{eq_zsup}).

\section{CONCLUSIONS}

By employing overlap fermions at two values of the lattice spacing in
the quenched approximation we tested whether or not the scalar
condensate $\Sigma$ obtained from a study of the behaviour of
$(a\mps)^2$ as a function of the quark mass is compatible with the one
previously extracted from a completely independent finite-size scaling
method.

We found this test to come out positive at leading order in quenched
$\chi$PT and postpone the discussion of higher-order effects to future
work \cite{cond:paperIV}. We also determined the pseudoscalar decay
constant, as well as the non-perturbatively renormalized strange quark
mass. Within the relatively large statistical errors, the agreement
with results obtained in the continuum limit using O($a$) improved
Wilson fermions \cite{mbar:pap3}, the overlap results of
ref. \cite{cond:boston} and also the Domain Wall formulation
\cite{dwf:RBC_spect} at non-zero lattice spacing is excellent. There
are strong indications that cutoff effects for overlap fermions are
small, but more statistics as well as a larger range of lattice
spacings is required to corroborate these findings.

\section*{ACKNOWLEDGMENTS}

We are grateful to Jochen Heitger and Rainer Sommer for permission to
use the ALPHA Collaboration data. LL thanks the INT at the University
of Washington for its hospitality and the DOE for partial support
during the completion of this work. This work was supported in part by
EU-TMR project FMRX-CT98-0169 and EU-IHP project HPRN-CT-2000-00145.

\end{document}